# Celebrating 30 Years of K-12 Educational Programming at Fermilab

M. Bardeen, Manager Education Office
*FNAL, Batavia, IL 60510, USA*

M. P. Cooke, DZero Experiment
*FNAL, Batavia, IL 60510, USA*

In 1980 Leon Lederman started Saturday Morning Physics with a handful of volunteer physicists, around 300 students and all the physics teachers who tagged along. Today Fermilab offers over 30 programs annually with help from 250 staff volunteers and 50 educators, and serves around 40,000 students and 2,500 teachers. Find out why we bother. Over the years we have learned to take advantage of opportunities and confront challenges to offer effective programs for teachers and students alike. We offer research experiences for secondary school teachers and high school students. We collaborate with educators to design and run programs that meet their needs and interests. Popular school programs include classroom presentations, experience-based field trips, and high school tours. Through our work in QuarkNet and I2U2, we make real particle physics data available to high school students in data-driven activities as well as masterclasses and e-Labs. Our professional development activities include a Teacher Resource Center and workshops where teachers participate in authentic learning experiences as their students would. We offer informal classes for kids and host events where children and adults enjoy the world of science. Our website hosts a wealth of online resources. Funded by the U.S. Department of Energy, the National Science Foundation and Fermilab Friends for Science Education, our programs reach out across Illinois, throughout the United States and even around the world. We will review the program portfolio and share comments from the volunteers and participants.

## 1. Introduction

Leon Lederman began his tenure as Fermilab's director in June 1979. With the inauguration of Saturday Morning Physics in fall 1980, Leon began another tenure as an advocate for K-12 science education. Why bother? Leon said, "Why not use the magnificence of Fermilab to dazzle (and capture) high school kids?" Saturday Morning Physics started with a handful of volunteer physicists and 300 high school students. When the first announcement went out, teachers asked to tag along with their students. Leon learned that the teachers did not know much about particle physics but were eager to learn, and he knew he wanted to offer a program for teachers too. Without money to do this, he encouraged a small group of volunteers to establish a non-profit organization, Friends of Fermilab, to support more programming. Friends offered the next program, the Summer Institute for Science Teachers that ran for 11 years. The rest is history! From this small beginning, Fermilab programs have matured to support the teaching and learning of science, technology, engineering and mathematics in the neighborhood and around the world.

## 2. Why Bother?

Professional development activities at Fermilab have a real impact on teaching and learning. Teachers apply new content knowledge and try new teaching practices in their classrooms. For example, Bob Grimm, Fremd High School, Palatine, Illinois, has "sprinkled" activities based on particle physics throughout his regular physics classes—data, computer simulations and anything kids would call "sweet" made it into his classroom. Pete Bruecken, Bettendorf High School, Bettendorf, Iowa, has "reloaded his curriculum" and tried teaching practices based on a renewed understanding of how physicists work. This has prompted class discussions about particle physics research and empowered students. Deborah Roudebush, Oakton High School, Vienna, Virginia, has woven the search for Higgs as a yearlong theme in her physics classes. Her students have remembered talking about Fermilab and CERN, researching the Higgs and the fact that their teacher helped build the ATLAS detector.

Offering programs for K-12 teachers and students is a win-win proposition. Physicists share their passion for science and inspire the next generation of scientists . . . and tomorrow's taxpayers. Anne Heavey, Fermilab Computing Division comments, "I love spending time with the kids . . . always trying to spark an interest." Teachers gain the respect of physicists, boosting their self-esteem as teachers and bring today's physics into their curriculum. Carol Baker, Alan Shepard High School, Palos Heights, Illinois, notices that her students are "transformed from high school science students to research scientists." Students find today's science inspiring and often get a chance to experience science in action. Ting Wu, Illinois Mathematics and Science Academy, Aurora, Illinois, has said, "We learned that mistakes are not dead ends, but simply stepping stones."

### 3. Volunteers

Technical staff volunteers—scientists, engineers, computing specialists, and technicians—keep Fermilab's K-12 programs on track. Michael Cooke is one of those exemplary volunteers. He began giving tours to college students after moving to Fermilab full-time. He soon added Ask-s-Scientist Q&A sessions with secondary student visitors to his schedule. Next in Michael's portfolio were classroom presentations. He is an expert at the *Charge! Electricity and Magnetism* demonstrations for grades 2 though 8, has developed presentation materials to accompany that demonstration kit, and helps train new volunteers to do the presentation. Mike worked with David Schmitz to develop a science show that they perform at the Family Open House and during Daughters and Sons to Work Day. In 2009 they were on stage at Millennium Park in Chicago to present a cryogenic demonstration as part of Science Chicago's Lab Fest.

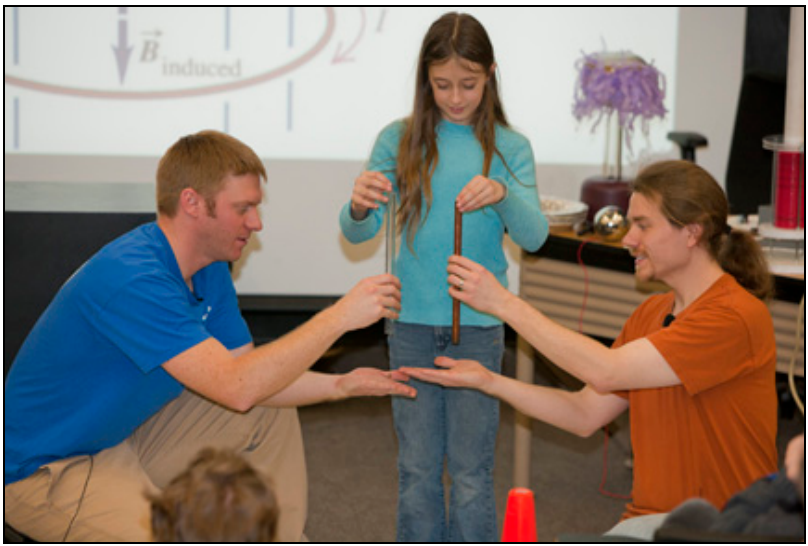

Figure 1: Mike Cooke (R) and David Schmitz (L) at Fermilab's Family Open House

### 4. Challenges

Over the years, we have confronted a number of challenges to make particle physics accessible for middle school and high school students. Finding the correct topical level is critical and not always easy. Simplifying may compromise precise explanations, and physicists can argue about how much should be left in or can be taken out. Teachers can help physicists determine the level. Additional resources that dig deeper into the topic help provide more complete explanations. Fitting content into the curriculum is also a problem because of the many pressures on school curricula. Standards, high-stakes testing, time and more infringe on a teacher's ability to incorporate new material. Providing a variety of activities from short "sprinkling" activities to full instructional units gives teachers an opportunity to choose what works at their school. Educational research shows that effective learning takes place when students actively participate in their learning and teachers facilitate rather than lecture. Figuring out how to engage young students with particle physics is not easy, and preparing materials of "publishable quality" usually requires the assistance from professionals in instructional design. Helping teachers move from "sage on the stage" to "guide on the side" may take sustained contact over several years.

### 5. Program Development

How we develop programs is just as important as the programs themselves. At Fermilab we begin by holding a needs assessment with educators to determine their needs and interests. When we find a match with Lab resources, we invite a few master teachers to help us develop and then conduct a program. Putting peers in leadership positions helps validate the program and provides opportunities for teachers to become "professional" by working outside their classrooms. Of course, participant feedback provides valuable information to improve programs.

Our experience has shown that activities that engage students in explorations of the world of particle physics are most effective. In the short term, we can create interest and awareness, in the longer term, understanding. We can also build lasting relationships that foster learning.

## 6. Programs and Resources

From the programs and resources, we will explore online resources, classroom presentations and the Family Open House.

Table 1: Current Fermilab K-12 Education Programs [1]

| **Research Experiences** | **Teacher Resource Center** | **Special Events for Kids and Families** |
|---|---|---|
| Academic Year High School Interns | Resource Collections | DUSEL Education Collaboration |
| QuarkNet Summer Research (Student/Teacher Teams) | Workshops | QuarkNet Masterclass |
| TARGET | Chem West | STEM Career Fair |
| TRAC | **Classroom Resources** | Wonders of Science |
| **Field Trips/High School Tours** | Classroom Presentations | Family Open House |
| Lederman Science Center | Data for Students | Family Outdoor Fair |
| Physics Science Experiences | I2U2 e-Labs & Science Investigations | **Professional Development for Teachers** |
| Beauty and Charm | Online Resources | Fermilab/U Chicago QuarkNet Center |
| Phriendly Physics | What is scientific research? | I2U2 Teacher Workshop |
| Prairie Science Experiences | **Classes for Kids** | Physics Experiences Teacher Workshops |
| Insects at Work in Our World | Prairie Rangers | Prairie Experiences Teacher Workshops |
| Prairie - Our Heartland | Saturday Morning Physics | Prairie Workshops |
| Particles and Prairies | Science Adventures | QuarkNet Boot Camp |
| Tours | Scout Programs | QuarkNet Outreach |
| | | Summer Secondary Science Institutes |

### 6.1. Online Resources

One of our newest efforts uses the theme of discovery science to present activities and resources for high school physics teachers and resources for high school students. We began this website to help people understand that behind the headlines of a new discovery are years of work by lots of people. The story line begins with the Standard Model as the current framework for our understanding of matter. While it explains much, it leaves many unanswered questions. Physicists build new instruments to address some of these testable questions. Particle physics research relies on indirect evidence to support claims. They check for statistical uncertainties and systematic uncertainties and whether other experiments are able to verify new discoveries. With this background, teachers and their students can follow along as physicists report new results.

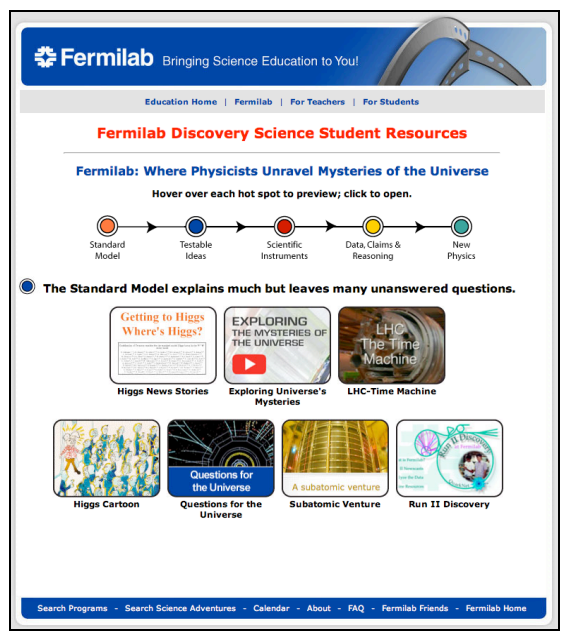

Figure 2: Discovery Science webpage for student resources [2]

Other online activities include a number of short investigations written by physics teachers with assistance from physicists. Examples include *Calculate the Mass of the Top Quark* based on DZero data; *the Mass of U.S. Pennies,* a study of histograms; the *Quark Workbench,* puzzle pieces that obey rules of the Standard Model limiting quark composition of bound states; and *Superconductivity Activities* in which students can experimentally discover the temperature dependence of resistance.

We have worked with the QuarkNet [3] and Interactions in Understanding the Universe (I2U2) [4] collaborations to provide online data from DZero, CDF, CMS and student cosmic ray detectors. CMS, Cosmic Ray and LIGO "e-Labs" explore the potential of using the Internet and distributed computing in high school classes and provide opportunities for students to organize and conduct authentic research and experience the environment of scientific collaborations. From start to finish e-Labs are student-led, teacher-guided projects. Students need only a web browser to access grid techniques employed by professional researchers. A project map with milestones allows *students* to set the research plan rather than follow a step-by-step process common in other online projects. Most importantly, e-Labs build the learning experience around the students' own questions and let them use the very tools that scientists use.

Figure 3: CMS e-Lab project map and milestone [5]

## 6.2. Family Open House

Our *Family Open House* is an afternoon of fun held in early spring. Younger children bring an adult with them to learn about the world of physics. From cryogenic shows to Ask-a-Scientist, from tours of the Linac or Magnet Factory to hands-on activities, hundreds of families are busy exploring Fermilab's physics. In 2010, we invited teams of high school students to create hands-on activities. Their physics carnival was a great hit and brought more girls than boys to Fermilab!

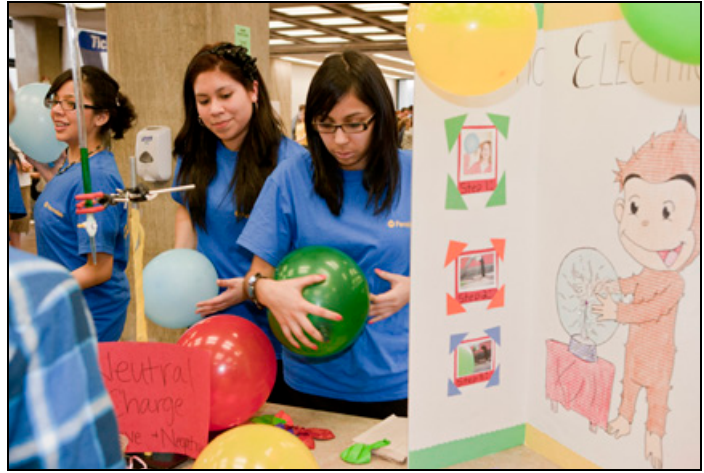

Figure 4: High school students prepare hands-on activity for the Family Open House.

### 6.3. Classroom Presentations

We also put on exciting presentations at elementary and middle schools near the Lab. Topics include the *Cryo Show*, *Charge! Electricity and Magnetism*, *Forces and Motion*, *Light and Color*, and more. *Space, Time and Einstein* is for high school students. The 90 scientists, engineers and computer specialists who are ready to drive a van of engaging demos owned by Fermilab Friends for Science Education [6] visit over 16,000 students each year.

## 7. Conclusion

Today, Fermilab offers over 35 K-12 programs. The Education Office has 14 staff members, and 25 docents lead tours and field trips. In FY10, participants included over 37,000 students and 2,500 teachers. As many as 50 educators worked as program developers and instructors; 250 members of the technical staff volunteered to help with programming. Thanks to Leon for starting down the road that has led to a vibrant program for K-12 teachers, students and families. *We can* use the magnificence of Fermilab to dazzle (and capture) kids of all ages!

### Acknowledgments

This work is supported in part by Fermilab, the U.S. Department of Energy, the National Science Foundation and various donors through Fermilab Friends for Science Education. The authors appreciate support and assistance from colleagues on the fabulous team that provides the K-12 educational programming at Fermilab. We also thank LaMargo Gill who edited this manuscript.

### References

[1] http://ed.fnal.gov
[2] http://ed.fnal.gov/projects/fnal
[3] http://quarknet.fnal.gov
[4] http://www.i2u2.org
[5] http://www18.i2u2.org/elab/cms/home/index.jsp
[6] http://ed.fnal.gov/ffse